**Table of Content graphics:**

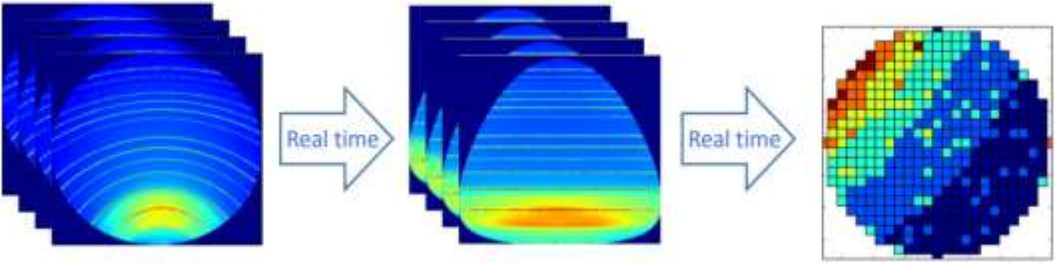

# On-the-fly Data Assessment for High Throughput X-ray Diffraction Measurement

*Fang Ren[1,*], Ronald Pandolfi[2], Douglas Van Campen[1], Alexander Hexemer[2], and Apurva Mehta[1, *]*

[1]Stanford Synchrotron Radiation Lightsource, SLAC National Accelerator Laboratory, Menlo Park, CA 94025, USA

[2]Advanced Light Source, Lawrence Berkeley National Laboratory, Berkeley, CA 94720, USA

KEYWORDS: high throughput, x-ray diffraction, data quality assessment, attribute extraction, on-the-fly

ABSTRACT: Investment in brighter sources and larger and faster detectors has accelerated the speed of data acquisition at national user facilities. The accelerated data acquisition offers many opportunities for discovery of new materials, but it also presents a daunting challenge. The rate of data acquisition far exceeds the current speed of data quality assessment, resulting in less than optimal data and data coverage, which in extreme cases forces recollection of data. Herein, we show how this challenge can be addressed through development of an approach that makes routine data assessment automatic and instantaneous. Through extracting and visualizing customized attributes in real time, data quality and coverage, as well as other scientifically relevant information contained in large datasets is highlighted. Deployment of such an approach not only improves the quality of data but also helps optimize usage of expensive characterization resources by prioritizing measurements of highest scientific impact. We anticipate our approach to become a starting point for a sophisticated decision-tree that optimizes data quality and maximizes scientific content in real time through automation. With these efforts to integrate more automation in data collection and analysis, we can truly take advantage of the accelerating speed of data acquisition.

INTRODUCTION

Advanced technology requires functional materials with tailored properties. Rational materials design based on prior materials science knowledge is appealing because it avoids the tedious trial-and-error of a routine material discovery cycle. Under the umbrella of the Materials Genome Initiative several large-scale efforts have been undertaken to combine high throughput computational material science with intelligent data mining to accelerate prediction of new materials[1-2]. However, detailed quantitative knowledge of structure-property-processing relationships is often still missing, especially for compositionally complex and non-equilibrium materials[3]. The predictions from these efforts must still be experimentally verified; and once verified, composition and processing conditions need to be tweaked for optimized performance. A recent estimate states that there are at least $2^{86}$ experimentally unexplored chemical systems[3]. The rapid parallel screening of a large number of samples provided by high throughput (HiTp) experimentation is, therefore, sorely needed[4].

In HiTp experimentation, syntheses and characterizations are coupled. HiTp synthesis, often described as "combinatorial material libraries", has the advantages of requiring small amounts of materials for each sample and the ability to synthesize a large number of compositionally permutated samples simultaneously[5-6]. X-ray diffraction (XRD) using an area detector is a rapid structural characterization technique; it can be easily paired with non-destructive compositional analysis techniques such as the x-ray fluorescence (XRF) measurements to build composition-property connections quickly[7]. Thanks to investment in brighter light sources, larger and faster detectors and the automation efforts, the data acquisition speed at national user facilities, such as at synchrotron-based HiTp XRD facilities, have caught up with HiTp material production[8-10]. For example, as of mid-2016, a ternary diagram with about 1300 compositionally varying samples can be mapped within 15 h at the HiTp XRD facility at Stanford Synchrotron Radiation Lightsource (SSRL).

However, the progress on the data acquisition creates new challenges for the analysis of data. The unmatched speed between data acquisition and data analysis severely hampers the wealth of new data to drive new scientific discoveries. This major short-coming was extensively explored in a 2015 DOE workshop and highlighted in the findings in a report titled, "Management, Visualization, and Analysis of Experimental and Observational Data (EOD) - The Convergence of Data and Computing"[11]. One of the major problems the workshop indicated is that, because the large amount of data acquired in a short period of time was not assessed as it was collected, less optimal data was often collected, which in extreme conditions forced repetition of measurements and thus wasted expensive characterization resources. In order to truly take advantage of the HiTp XRD data acquisition capability, we developed several new algorithms and software tools for assessing the data quality as soon as they are collected. The design and deployment of these data assessment tools allows the optimal use of the characterization resources, and allows the data to be ready for more sophisticated analysis. Developing such data assessment tools is the first part of our efforts to meet the goals set by the DOE workshop.

The first step in our on-the-fly data assessment approach is to reduce raw XRD data into physically meaningful formats as soon as it is collected. Next, physical information in the form of attributes, which represent various physical properties or characterization control parameters, are extracted from the reduced data on-the-fly on a generic desktop machine. We strived to keep analysis close to the data source in both time and space to minimize the amount of data moved across networks. These attributes are then used to assess data quality for both individual (local data) and a collection of diffraction patterns (global data). Because this information is close to the data source, it can enable optimization of data quality on the fly as well as allows scientists to strategize data collection plans and prioritize experiments in real time. We anticipate our result to be a starting point to build a sophisticated decision-tree to automate data quality improvement. Our development also allows more advanced machine learning-based analysis, such as automated phase identification[12-15], to be running in parallel to data acquisition in the future.

RESULTS AND DISCUSSIONS

The data acquisition speed of XRD analysis is significantly boosted by using large 2D detectors. Instead of scanning diffraction intensity by varying Bragg angles, 2D detectors record complete XRD data in a single exposure, usually in less than a minute. The XRD data produced by a 2D detector is a 2D image. The pre-requisite for rapid data assessment is to automatically reduce the raw 2D image into scientifically relevant formats as soon as it is collected.

I: *On-the-fly data reduction*

An example of a raw XRD image for Lanthanum hexaboride ($LaB_6$) is shown in Figure 1(a), and the Miller indices for XRD "arcs" are shown on the image. $LaB_6$ is commonly used as a standard material to calibrate the detector geometry. Such a raw XRD image is first mathematically converted into a calibrated XRD image in the diffraction coordinate system, the Q-γ plot (Figure 1(b)), based on the knowledge of detector placement with respect to the position and direction of the incident x-ray beam. The corresponding Miller indices are also shown on the plot. In Q-γ plots, the x-axis, "γ", measures the Azimuthal angle between the diffracted beam and the vertical plane containing the incident beam. "Q", the y-axis, is the scattering vector defined as the momentum transfer between the diffracted beam ($Q_d$) and incident beam ($Q_i$), as in Figure 1(d). In a Q-γ plot, if calibrated correctly, the scattering arcs become straight lines unless there is deviatoric strain in the sample. The 1D spectrum is produced from the corresponding Q-γ plot by averaging the intensity along "γ" (the left y axis in Figure 1(c)). Because Q is related to Bragg angle (2θ) by the relationship Q = 4*π*sin(θ)/λ, with λ the x-ray energy, a more traditional 1D spectrum (intensity vs. 2θ) can then be generated as shown in Figure 1(c) with the right y axis. Several existing software packages can reduce raw images to 1D spectra, for example NIKA[16], WxDiff, GSAS II[17] and Fit2D[18]. However, these software packages were written mainly for single image analysis with some batch processing functionality, but not optimized for unsupervised on-the-fly analysis. To build a flexible workflow with on-the-fly capability, we took advantage of a recently developed GPU-integrated pyFAI library[19] and adopted the "tilt-rotation" geometry for XRD data reduction, with polarization correction[20]. Our software looks for a newly created raw image in a folder, transforms it into a Q-γ plot and 1D spectrum, and then performs other experiment-specific operations. We believe that our software is applicable to any small or wide angle scattering pattern collected on a monolithic 2D detector. The current data reduction speed is of the order of 0.1 s per image on a generic desktop machine at SSRL Beamline 1-5.

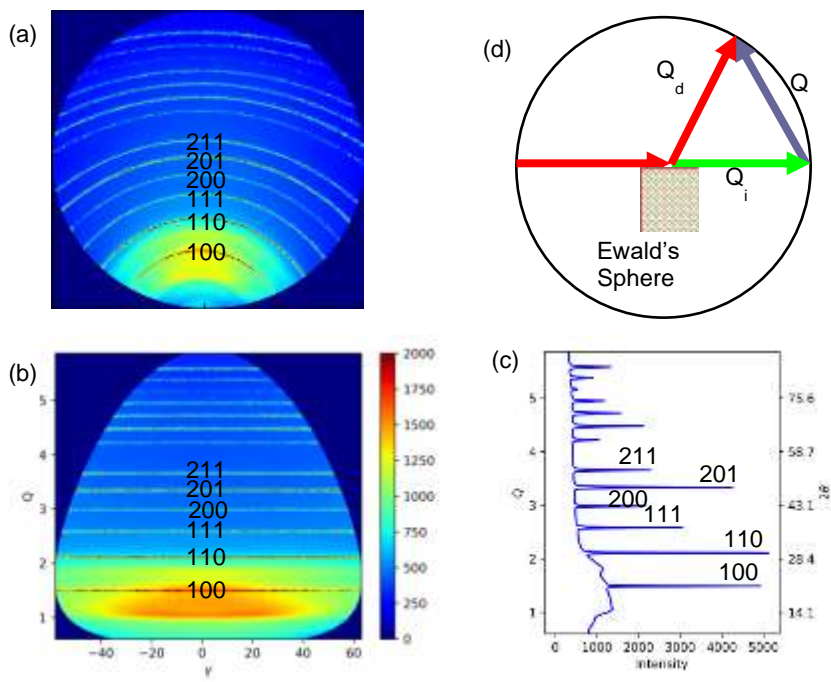

Figure 1: Data reduction of the XRD data for Lanthanum hexaboride. (a) raw image, (b) Q-γ plot, (c) 1D spectrum in Q space (left y axis) and Bragg angle space with copper K α x-ray source (right y axis), (d) definition of momentum transfer "Q".

II: *HiTp XRD data structure*

Now that the data has been reduced to scientifically meaningful formats, analysis can progress with extraction of scientifically relevant information. Extraction of this information requires understanding of different raw data streams and the context of each stream in the overall experimental design. Below in Figure 2, we illustrate this multi-layered tree structure of HiTp data from a combinatorial material library (combi library or combi wafer). The top layer, i.e. the root of the tree, describes the alloy system under investigation. The next layer in the tree describes how the alloy system is broken down into several combi libraries. It also describes the processing conditions used for synthesizing each of the libraries. This layer, therefore, captures how the experimental plan is executed. The next layer describes individual samples, i.e., individual samples in the combi libraries. In the case of discrete libraries where the samples are physically separated, sample definition is self-evident. However, in the case of continuous compositional spread, the samples are defined artificially by the characterization scan grids. Information contained in this layer becomes especially critical when two different characterization streams need to be compared or combined. The next layer describes the various types of characterizations included in the experimental plan, followed by several layers of data. This first layer of data contains raw data (primary data), in the forms of images and spectra from various characterization techniques, as well as metadata describing characterization conditions necessary to process the raw data. The layers below contain various "data derived products" (secondary data), which is the focus of this paper. We broadly separate the "data derived products" into reduced data, for example Q-$\gamma$ images and 1D spectra, and attributes extracted from them (often as scalars, but sometimes as a 1-D vector).

In the examples shown in this article, each sample was characterized by XRD and XRF. The metadata associated with raw data records exposure time, sample location, sample height, x-ray energy and so on. In the discussion below we will talk about either local data, i.e., data (XRD or XRF data) for a single sample or global data, i.e., data ensemble (or dataset) collected for a combinatorial wafer (combi wafer) or a ternary system.

In our approach, the goals of data assessment fall under three categories described below. The first category uses local data, whereas the second two categories use global data.

1) To assess the data quality of local data (single spectrum). For example, the situations that the sample is under-exposed, yielding low signal-to-noise ratio, or over-exposed, causing signal saturation and peak flattening, should be avoided.

2) To assess the data quality of global data in order to detect an experimental incident. For example, data collection should be aborted immediately when a machine error occurs or an *in-situ* experiment is complete.

3) To assess the data quality of global data in order to optimize the usage of characterization resources according to the scientific values of each sample. For example, samples with flaws or are not scientifically interesting should not use the same amount of characterization time as those with higher scientific impact.

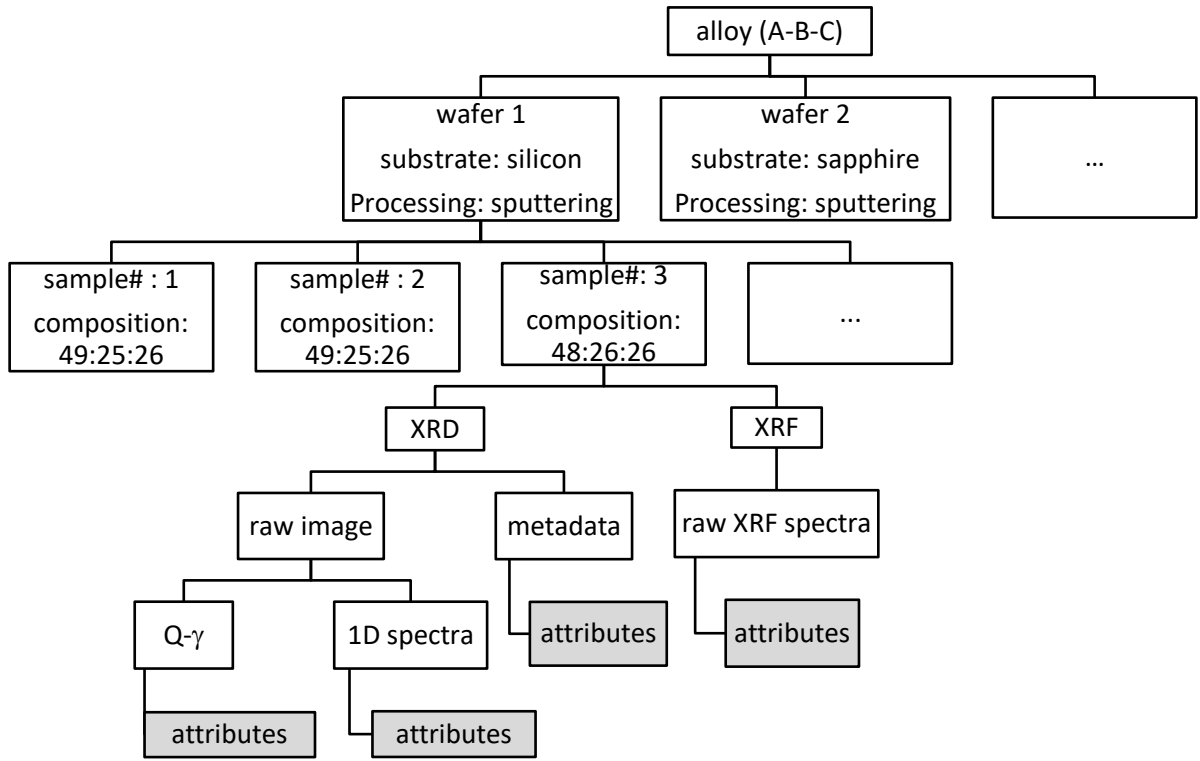

Figure 2: HiTp data structure.

To achieve the goals of data assessment in the three categories discussed above, we extract scientific information that can help determine data qualities from not only the reduced XRD and XRF data, but also from metadata associated with them. This quantified information is represented by defining various "attributes". In the section below we illustrate our approach by using some common attributes relevant to combi libraries.

III: *Attribute extraction*

1) Data quality of local data

*Signal-to-noise ratio*

The most direct way to measure the data quality of a single spectrum is to monitor the signal-to-noise ratio (SNR). Figure 3 compares five spectra with different noise levels. The SNR is calculated by dividing the power of signal (sum of square of signal intensities in Figure 3(a)) by the power of noise (sum of square of noise intensities in Figure 3(b)), and the noise spectrum is generated by subtracting the smoothed spectrum from the original spectrum (see the Experimental Methods section for more details). Using this method, Spectrum 1 (blue) has an SNR of 66.9 dB, whereas Spectrum 3 (red) has an SNR of 50.3 dB. The SNRs of Spectrum 1 (blue) and the Spectrum 3 (red) are both sufficiently high to not mask any diffraction features. With limited characterization resources, collection of data with a high SNR as Spectrum 1 is a waste of time. On the other hand, the noise in Spectrum 4 (cyan) is significant (SNR = 21.7 dB) and some of the peaks are masked by noise. By increasing the exposure time by 10 times, the SNR of Spectrum 5 (magenta) increases to 31.6, and now it becomes easier to separate most of the peaks from noise.

The challenge in determining SNR using this method is the accurate estimation of noise. If the frequency of peaks in a spectrum is smaller than the dominant noise frequency, then smoothing the spectrum and subtracting it from the raw spectrum gives a reasonably accurate measure of the noise. However, when the signal peak widths are comparable or smaller than noise peaks, like in Spectrum 2 (green), the method described above to estimate noise power directly from noise spectrum (Figure 3(b), green curve) will overestimate the noise as it incorporates some of the peak features in the noise spectrum. The SNR estimated for Spectrum 2 (green) using this method is 27.7, which is smaller than that of Spectrum 5 (magenta), as in Figure 3(d). An SNR of 27.7 does not represent the true quality of the Spectrum 2, in which the peaks are readily distinguishable from noise. Whereas in Spectrum 5, which has an SNR of 31.6, the noise and peaks are difficult to separate, for example the features around Q = 3.45. Therefore, a robust method for estimating noise is needed.

A better way to characterize the noise is to examine noise distribution and to fit the histogram using Gaussian functions (Figure 3($c_1$) – ($c_5$)). Because the number of signal peaks is much smaller than the number of noise peaks, the contribution from signal peaks in the histogram is small enough to be ignored. This is illustrated in Figure 3($c_1$) and ($c_2$); even if Spectrum 2 (green) has extra peaks outside the main distribution (appearing as a tail on the high end of the distribution), the Gaussian fitting will treat those peaks as outliers and produce a truer estimation of the noise.

We are interested in finding a relationship between a statistical value of the noise distribution, for example the full width at half maximum (FWHM) and SNR. A plot of SNR (dB) against log(1/Gaussian FWHM of the noise distribution) produces a strongly correlated linear relationship (Figure 3(d)), with a positive correlation coefficient of 0.993, suggesting that log(1/FWHM) is a reasonable measure of SNR. In our experience, acceptable quality XRD spectra have SNRs of at least 30. Estimation of the noise as the FWHM of the noise distribution produces SNR for Spectrum 2, which is similar to that for Spectrum 1, indicating that both Spectrum 1 and 2 are of high quality.

On-the-fly determination of SNR can thus enable a strategy that counts longer in the parameterized regions where data is noisier at the expense of regions where the signal is stronger for an optimum use of characterization time without losing information contained in a dataset.

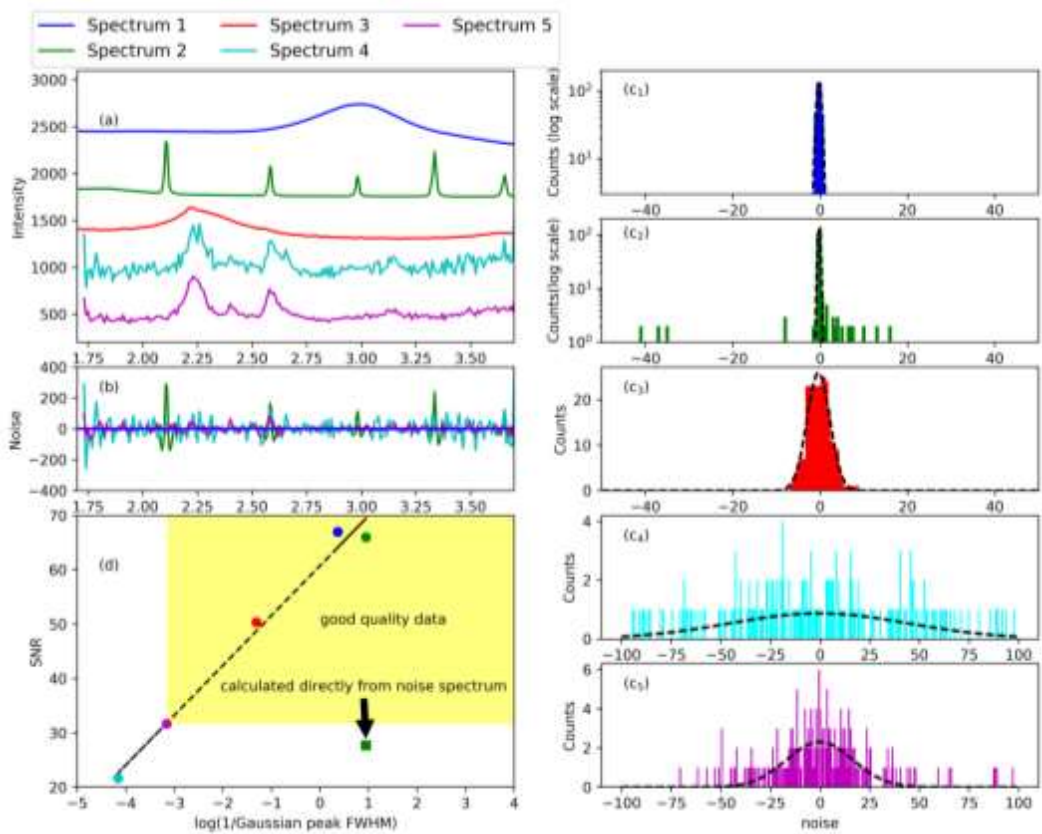

Figure 3: Multiple plots show how various spectra are analyzed to assess their quality level: (a) raw spectra; (b) noise spectra; ($c_1$)-($c_5$) noise histograms with Gaussian fit; (d) linear relationship between SNR and log(1/FWHM). Blue color shows Spectrum 1; green shows Spectrum 2; red shows Spectrum 3; cyan shows Spectrum 4; and magenta shows Spectrum 5, with 10x exposure time of Spectrum 4.

*Sample height*

Another useful single spectrum attribute for XRD data from combi libraries is the precise location of the incident beam on the library. Most HiTp samples are supported on flat substrates. To avoid signals from substrates, the HiTp XRD experiments at SSRL are performed in the grazing incidence reflection geometry (grazing incidence angle is usually between 1 to 8 degrees and most commonly at 3 degrees). At an incident angle of 3 degrees, 50 µm of change in sample height will cause the x-ray beam to displace by 1 mm on the sample (See the illustration S1 in the Supporting Information). Because the compositions vary by location on combi libraries, the sample height must be precisely adjusted before recording each XRD pattern. In the HiTp experimental setup at SSRL, the sample height is constantly monitored by a laser distance finder and recorded as metadata. The sample height can either be used for a real-time correction or for composition correction after data collection.

Global data assessment for HiTp data is important because it is possible that even though each of the XRD spectra is of high quality, the data may not have much value in the context of the scientific goals for the experiment. To assess the quality of global data, either single spectrum attributes are compared with their neighbors, or multi-spectra attributes are extracted and visualized to capture changes in XRD spectra from point to point. Because HiTp samples are often synthesized with composition gradients, producing gradual property gradients, one must be cautious when there is a sudden change in any property between neighbors. Such abrupt changes often indicate occurrence of experimental mishaps.

2) Experimental incidents

*XRF elemental channel*

XRF analysis is a direct way to monitor compositional changes; therefore, XRF signal is recorded simultaneously with XRD data to generate mappings of pre-defined XRF channels. Figure 4 presents two examples of machine errors that were suggested by XRF mapping. The first plot shows a mishap when part of the sample holder was blocking the x-rays from the sample, resulting in a sudden and unexpected change of XRF signal on the left side of the wafer. In Figure 4(b), the "jagged" XRF horizontal gradient across the wafer was likely due to either wafer drifting or a warped wafer surface. Detection of such experimental incidents in real time is invaluable as it allows the experimenter to stop the experiments that have gone awry and correct the error immediately. Monitoring property changes with real-time visualization for *in-situ* experiments, in which case axes in visualization are replaced by temperature or time, are also useful. The other attributes that can be used for this application will be introduced in the following section.

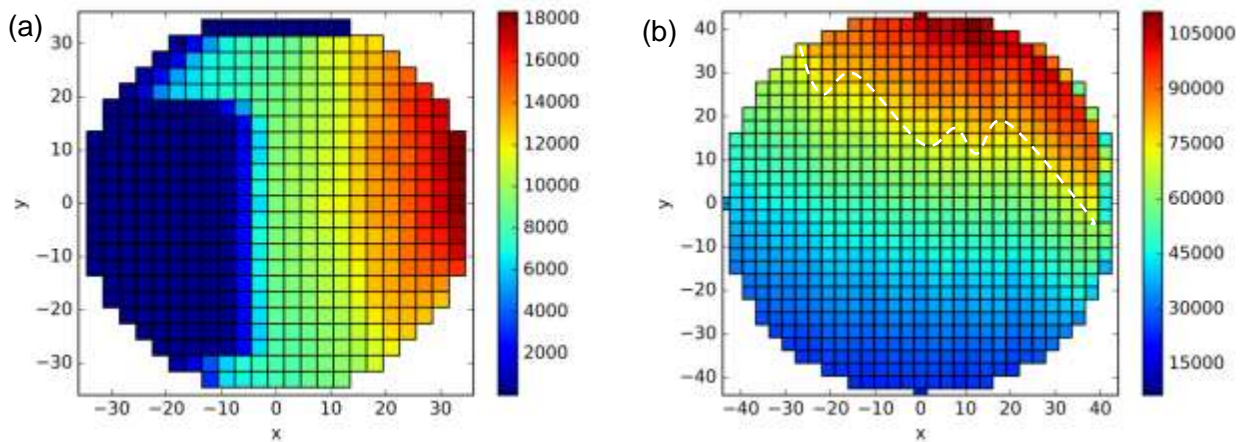

Figure 4: Two examples of bad quality data: (a) The wafer (25x25 mesh grids in a circle, 441 pads in total) was partially blocked by the sample holder. (b) The wafer (31x31 mesh grids in a circle, 709 pads in total) was drifting or warped. The white dotted line show abnormal XRF gradient of the wafer.

3) Prioritizing interesting samples

The second part of global data quality assessment is how to use the limited characterization resource to yield maximum scientific values. Guideline here is to spend more time and efforts on measurements which have highest impact on validating a scientific hypothesis or resulting in new discoveries. An effective strategy can be to perform a sparse data sampling across the whole wafer to rapidly locate regions that show interesting properties, and then to increase the data density on these regions of interest on the second pass. For this strategy to be effective, the results of the first screen must be available almost as soon as the measurements are completed, with minimum human intervention and low computational cost. Below, we illustrate such strategies for three types of sample characteristics - crystallinity, texture, and phase boundaries. In the end, we will present a case study using these methods.

*Crystallinity*

Commonly, degree of crystallinity is inferred from the full width at half maximum (FWHM) of XRD peaks. Because the integrated area of a powder diffraction peak (under kinematic conditions) is invariant of crystallinity, XRD peaks of more crystalline samples are not only narrower (with smaller FWHM) but also more intense (with larger peak intensity). Therefore, a ratio of the maximum intensity ($I_{max}$) to average intensity ($I_{ave}$) is also a good measure of crystallinity, and it is computationally easier and faster to extract from 1D XRD spectra than widths of diffraction peaks. Although technically, the two intensity values can be extracted directly from raw images, the validity of this method relies strongly on the algorithm to remove detector image "zingers" (artificial bright spots on the detector) which produce artificially high intensities. Because the intensity of every point of a 1D spectrum is averaged over hundreds of pixels across "$\gamma$", such artificial zingers in a 1D spectrum are mostly averaged out. The average intensity of a 1D spectrum is dominated by the diffuse and inelastic scattering, while the maximum intensity is almost always the intensity of the most intense peaks. Figure 5 compares two 1D spectra with different $I_{max}/I_{ave}$ values. Sample A has a sharp peak and is crystalline, while Sample B has a broad low peak and is most likely amorphous, with some likelihood of being nanocrystalline. This illustrates that with appropriately calibrated thresholds, the degree of crystallinity of each sample can be determined quickly by monitoring $I_{max}/I_{ave}$. In our analysis of the data collected at SSRL BL 1-5, the $I_{max}/I_{ave}$ is plotted on a log scale with a set maximum of 1.4 and a minimum of 0.2. Through cross-validation with peak width analysis, the amorphous-crystalline boundary is found to be around $I_{max}/I_{ave}$ = 0.6. These values can be adjusted accordingly for data collected with other instruments.

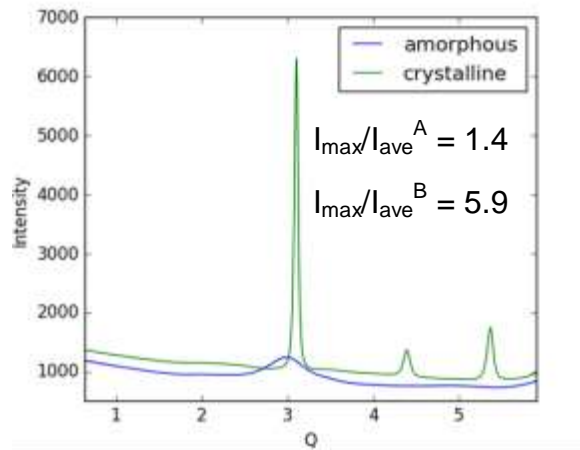

Figure 5: 1D spectra of Sample A (the green curve) and Sample B (the blue curve).

The attribute $I_{max}/I_{ave}$ can also be used to monitor the progress of in-situ experiments, for example, to facilitate the study of the thermal stability of an amorphous material. When the temperature is above the crystallization temperature, $I_{max}/I_{ave}$ is expected to increase dramatically. If a proper threshold is chosen, the scientist can be informed when the crystallization starts and decide whether to stop the experiment. Such automated monitoring is particularly invaluable for experiments that operate 24 h continuously.

*Crystallographic texture*

Many materials have large crystallographic anisotropy. There are sometimes advantages if these highly anisotropic materials are synthesized with a preferred crystallographic orientation. Locating materials with the desired crystallographic texture corresponding to some processing conditions is a valuable HiTp search. In powder diffraction, if the crystallites are randomly oriented, the XRD arcs in 2D image will be continuous, as in Figure 1. However, if the crystallites stack with preferred orientations, the continuous diffraction rings break up into discrete XRD arcs, as shown in Figure 6(a). Note that traditional 1D spectra (Figure 1(b)) which were created by integrating the intensity across the γ in Q- γ plots lack the texture information.

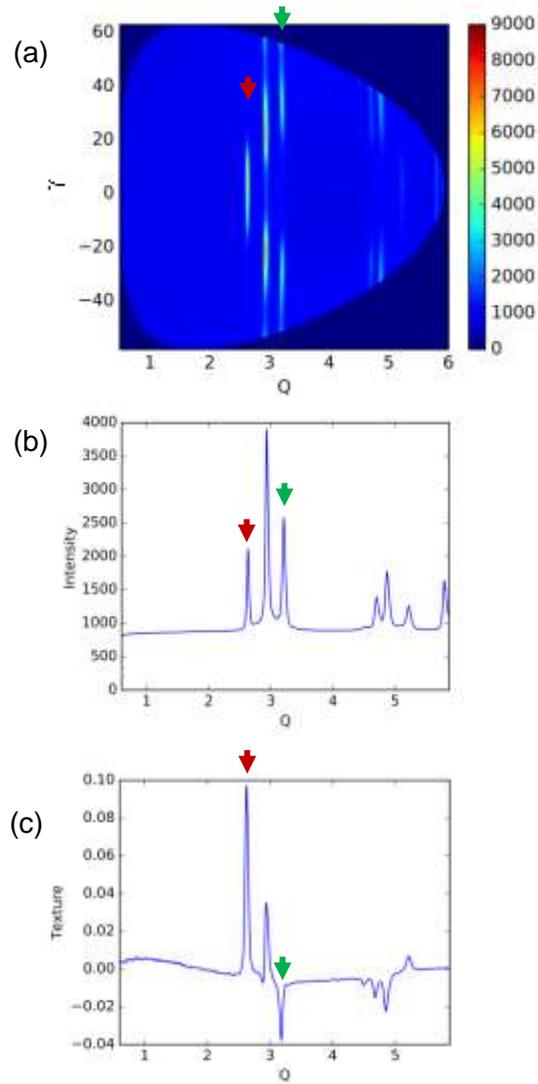

Figure 6: (a) Q- γ plot, (b) 1D spectrum, (c) texture spectrum of a textured sample. The red and blue arrows show different crystalline orientations.

A texture analysis algorithm was developed to monitor the crystalline texture in the sample. Note that to precisely measure the texture, some corrections[20] must be applied to the diffraction intensities. The texture analysis method introduced here is for flagging the textured samples that need extensive texture analysis. The as-defined texture compares azimuthally weighted peak intensity with the azimuthally averaged intensity. If the XRD arcs are continuous, the azimuthally weighted and azimuthally averaged intensities will be the same and therefore texture will equal zero. When the intensity is higher at γ close to 0 (resulting from out-of-plane preferred orientation), texture is larger than 0, denoted by the red arrow in Figure 6. When the intensity is higher at higher γ angle (resulting from in-plane preferred orientation), texture is smaller than 0, denoted by the green arrow in Figure 6. After generating a textured 1D spectrum, a texture_sum attribute is calculated by summing the square of all the texture values along Q in the texture spectrum and normalized by the number of Qs. The texture_sum can then be mapped on a log scale, with a maximum -10.3 of and a minimum of -11.1 in our examples, which can be adjusted for other instruments. Note that for an ideally non-textured sample texture_sum is zero, but a real XRD data with finite noise will show a small apparent texture_sum.

*Phase boundary – by peak numbers and nearest-neighbor distance*

Identification of phase boundaries is the essential step in discovering new materials in HiTp searches for combi libraries. An optimized search for new materials on a combi library should devote a larger proportion of the characterization resources on phase transition regions to identify the exact location of the boundary, while lower data density can be used to cover the region between boundaries where the structure of the material changes gradually. There has been great progress on developing phase identification algorithms, as mentioned in the introduction section. The goal of the on-the-fly analysis approach discussed in this article is not to attempt to solve the challenge of identifying phases of any materials in real time. Our goal is to ensure the data density is sufficiently high in regions where diffraction patterns are changing rapidly, so that more advanced phase identification algorithms can run reliably and with high accuracy.

Materials belonging to a single phase will have the same number of peaks in XRD spectra, though the peak positions may shift due to solid solution effects. Specifically, if crystalline phase A has 3 peaks, and crystalline phase B has 2 peaks, within the phase transition region (A→B), if there is an observable two-phase region, there will be temporarily 5 peaks because both A and B phases will be detected. The transition from phase A to B, thus, either goes from 3 to 5 and to 2 XRD peaks, or goes from 3 to 2 XRD peaks directly, depending on whether there is an observable two-phase region in between. Another scenario is a phase gradually loses crystallinity. In such a case, the peak number will monotonically decrease as the phase losing crystallinity. Therefore, a simple peak number map can locate the phase change and multi-phase regions so that the assessment of data density in these regions is possible.

Another method to monitor the phase changes is through an attribute that directly measures the distance between the nearest neighbors. The two neighbors in the upper stream of the scanning order (as shown in Figure 7) were chosen so that this method can run on-the-fly. While there are many dissimilarity matrices for identifying phase boundaries[21], the cosine dissimilarity matrix was chosen to measure the distances between XRD spectra because its performance in the presence of both peak height change and peak shifting when the magnitude of peak shifting is unknown[22]. This attribute has sensitivity to both peak shifting due to solid-solution effects as well as emergence or disappearance of peaks due to phase transformation. Therefore, this attribute is more universal than the peak number attribute.

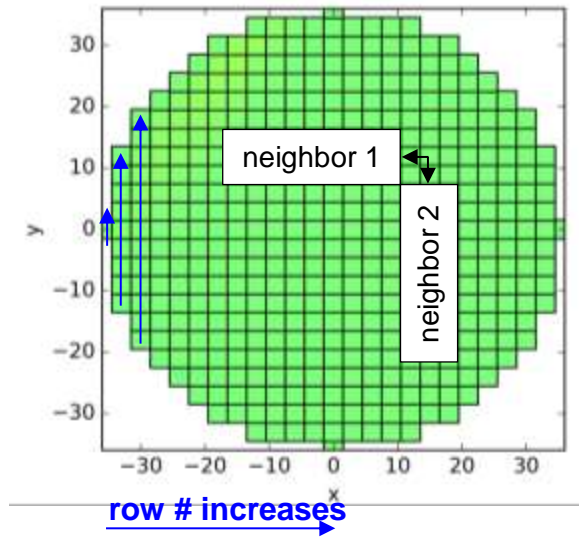

Figure 7: The relative locations of the sample-of-interest and its two nearest neighbors in the upper stream. The scanning order is shown by the blue arrows.

*Case study*

Below we demonstrate the on-the-fly attribute extraction methods discussed above coupled with visualization tools to illustrate how they operate with a single combi library. The elemental information from the combi library was omitted from the plots to avoid conflict of interest. Figure 8 displays four attribute maps for this dataset: crystallinity ($I_{max}/I_{ave}$) map, texture_sum map, peak number map, and nearest-neighbor distance map.

In the crystallinity map, the blue region is amorphous and the red region is crystalline. In a search for amorphous materials, locating the blue region associated with amorphous materials (Region 3 in Figure 8(a)) and the boundary between the amorphous and the crystalline regions (Region 2 in Figure 8(a)) is the primary focus of the experiment. In the texture_sum map, the blue region indicates untextured samples and the red indicates textured. If an experiment is performed to compare the substrate effect on nucleation film orientation, the samples in Region 1 (in Figure 8(b) are more likely to be influenced by the substrate than the other regions. In the peak number and nearest-neighbor distance maps, the regions where peak numbers and nearest-neighbor distance changes slowly, for example Region 3, can be covered with a low data density; whereas high data density (and characterization resources) can be devoted to regions with rapidly changing XRD patterns (Region 1 and 2). The guideline here is to use a higher data density on regions that have highest impact on validating a scientific hypothesis or resulting in a new discovery, depending on a particular application.

Based on the four attribute maps of the dataset shown in Figure 8, the researcher would know that Region 1 has crystalline and textured phases, 6-8 XRD peaks, and the spectra in this region change rapidly. The samples in Region 3 of the wafer are amorphous and untextured, and their peak numbers and nearest-neighbor distances do not change much. Region 2 captures the transition from Region 1 to Region 3, and thus a higher data density is recommended to precisely locate the transition boundaries. Depending on the scientific goals of the experiment, the research can decide whether the data density in Region 1 and 3 needs to be adjusted and additional data should be collected before the library is removed from the instrument.

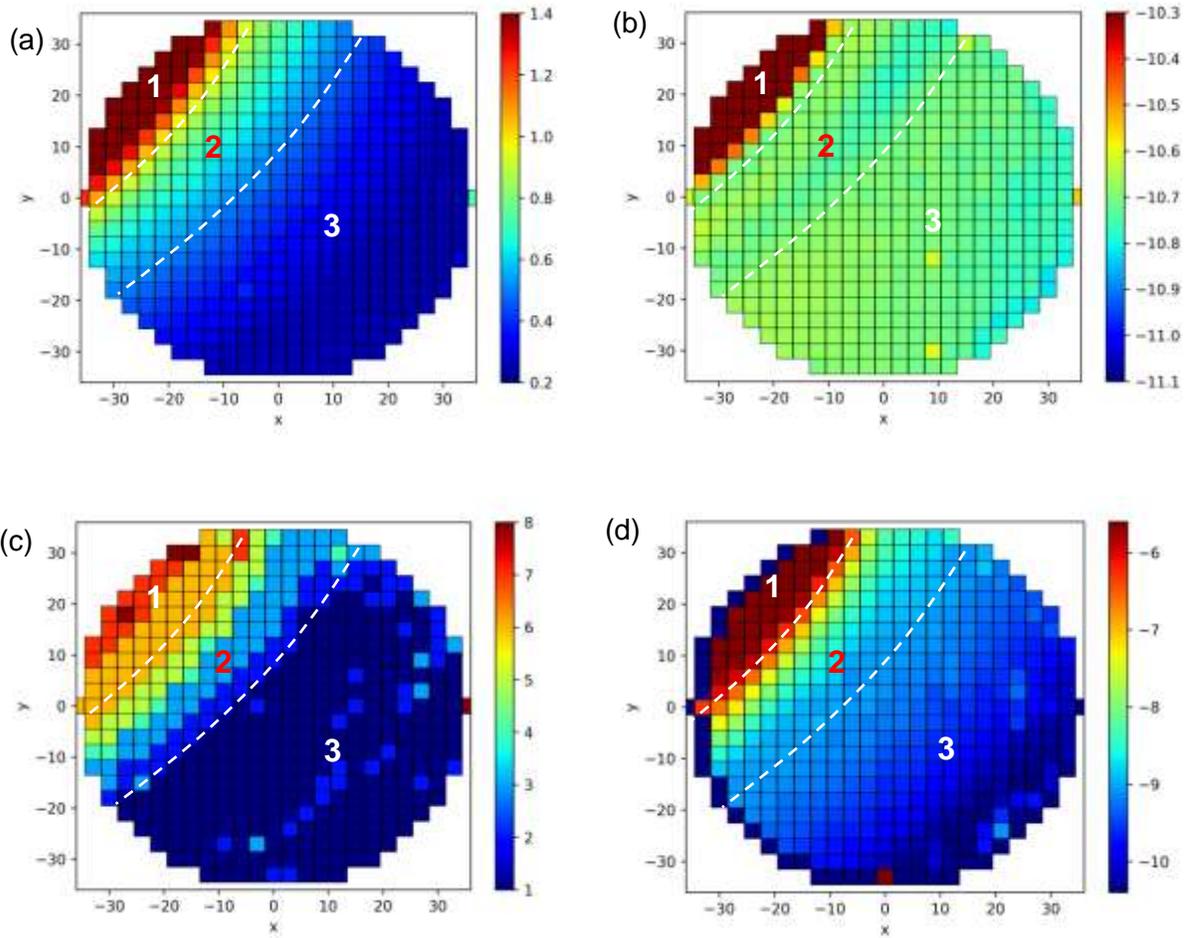

Figure 8: (a) $I_{max}/I_{ave}$ map of the wafer in log scale. (b) texture_sum map of the in log scale. (c) peak number map (d). Nearest-neighbor distance map in log scale. The wafer was mapped with a 25x25 mesh in a circle, with 441 pads in total. Two dotted curves separate the wafer into three regions: Region 1, Region 2, and Region 3.

CONCLUSION

The implementation of algorithms that enable on-the-fly assessment of high throughput data serves three critical functions. The first function of the data assessment is to ensure that the local data is of a desired quality. The second critical function is to assess the global data to detect experimental incidents or mishaps. The third function is to quickly detect regions of interest and trends in the global data and to assess whether those regions are covered with sufficient data density. Our goal is to provide assessment of data quality and coverage in real time so that scientists can decide whether the data quality is sufficient to address the problem that the experiment is designed for, and for more sophisticated analysis. These attributes also provide actionable information that scientists can use to enable strategic experimental plans towards experiments with the highest impacts. Intuitive visualization of the attributes is often as important as the design and extraction of attributes, as these visualizations are how actionable information is presented. The development and integration of these tools into an intuitive visualization platform that is made available to scientists during data collection has the potential to save hours and often days of expensive instrument time and human labor within each experiment cycle.

EXPERIMENTAL METHODS

The data presented in this paper are collected at SSRL beamline 1-5. The illustration of the scattering geometry is shown in a recent research paper[8]. The 2D detector used is MarCCD (2048-pixel*2048-pixel, 79 µm) from Rayonix, LLC. The raw images are calibrated using geometry parameters: sample to detector distance, detector tilting angle, rotation angle, and beam center on the detector ($x_{beam}$, $y_{beam}$). The geometry is shown in illustration S2 (Supporting Information).

*Signal to noise ratio*

SNR is calculated as the ratio of power of the raw signal and the power of the noise, as shown below[23].

$$SNR = \frac{P_{signal}}{P_{noise}}$$

$$P_{noise} = \frac{1}{N}\sum_{k=0}^{N}|noise(k)|^2$$

$$P_{signal} = \frac{1}{N}\sum_{k=0}^{N}|signal(k)|^2$$

To express the SNR in decibels,

$$SNR(dB) = 10\log_{10} SNR$$

The noise can be directly measured from instruments or estimated by subtracting the smoothed signal (using a Savitzky-Golay filter with a window size of 15) from the raw input signal. Ideally, the window size is chosen so that the noise is effectively captured while minimizing effect on signal features, which can be proven if the noise has a normal distribution.

$$noise(k) = signal(k) - SG\_smooth(signal(k))$$

However, with the current window size, the filter will not be able to differentiate noise peaks and crystalline peaks that have similar peak shapes. Therefore, the histogram of noise is plotted, and the distribution is fitted using a Gaussian function. The full width at half maximum (FWHM) of the fitted Gaussian peak is recorded to represent SNR.

$$SNR(dB) = a \times \log\left(\frac{1}{FWHM}\right) + b$$

a, b are constants.

*Texture analysis*

The definition of the crystallographic texture is shown below. In a Q-γ plot, at each q = $q_j$,

$$\text{intensity\_sum}_j = \sum_{\gamma_i}^{q=q_j} I_i$$

$$\text{weighed\_intensity\_sum}_j = \sum_{\gamma_i}^{q=q_j} I_i \times \cos(\gamma_i)$$

$$\text{count}_j = \sum_{\gamma_i}^{q=q_j} 1$$

$$\text{weighed\_count}_j = \sum_{\gamma_i}^{q=q_j} \cos(y_i)$$

$$\text{texture}_j = \left(\frac{\text{weighed\_intnsity\_sum}_j}{\text{weighed\_count}_j}\right) / \left(\frac{\text{intensity\_sum}_j}{\text{count}_j}\right) - 1$$

$$\text{texture\_sum} = \frac{\sum_Q^j (\text{texture}_j)^2}{\# \ of \ Q}$$

*Peak detection*

A continuous wavelet transform (CWT)-based peak detection is implemented by adopting a python library (scipy.signal.cwt), with a Ricker wavelet kernel[24]. Peaks are identified from the CWT convolution space by the ridge-line approach implemented in scipy. This algorithm takes into consideration characteristics of XRD peak shapes, providing robust peak detection on overlapping peaks and effectively separating signal from noise[25]. The parameters in CWT were chosen to favor false positives. After the "peaks" are found, a simple second-order finite differences threshold filter is applied to provide more precise peak detection results.

*Nearest-neighbor distance analysis*

The nearest neighbor distance attribute was calculated as a sum of cosine distances from the two nearest neighbors, one to the left and one below. If any of the neighbors is missing in the physical space, the distance is recorded as zero.

ASSOCIATED CONTENT

*Supporting Information.*

A file with two supporting figures is available online.

The python code is available to download through https://github.com/fang-ren/on_the_fly_assessment

AUTHOR INFORMATION

Corresponding Author(s)

*mehta@slac.stanford.edu, *fangren@slac.stanford.edu

Author Contributions

The methods included in this manuscript were developed by F. Ren and A. Mehta; F. Ren and R. Pandolfi developed the algorithms and implemented them so that they can run on a generic computer with sufficient speeds for real time analysis. F. Ren and A. Mehta collected and processed the x-ray diffraction and fluorescence data, with extensive hardware support from D.

Van Campen. F. Ren and A. Mehta wrote the manuscript, with the contribution of all authors. All authors have given approval to the final version of the manuscript.

The authors declare no competing financial interest.

Funding Sources

United States Department of Energy Office of Science, Office of Energy Efficiency and Renewal Energy and Advance Science Computing Research.

ACKNOWLEDGMENT

The authors acknowledge T. Dunn, S. Belopolskiy, V. Borzenets, and A. Maciel for their support at Beamline 1-5 and the Development Laboratory. The authors also acknowledge T. Duong for his early development on the nearest-neighbor distance analysis, Y. Liu for insightful discussion on SNR determination and B Ruiz-Yi from The University of South Carolina for providing an example of noisy data used in Figure 3. This work and the use of Stanford Synchrotron Lightsource (SSRL) were supported by United States Department of Energy, under the contract DE-AC02-76SF00515 and the Advanced Light Source under contract DE-AC02-05CH11231. This work was also partially supported by the Center for Advanced Mathematics for Energy Research Applications (CAMERA). R. Pandolfi and A. Hexemer's work was partially supported by A. Hexemer's Early Career Award from the US Department of Energy (DoE) and LBNL LDRD "TReXS".

ABBREVIATIONS

HiTp, High throughput; XRD, X-ray diffraction; combi libraries, combinatorial composition libraries.

application to light absorber discovery in the V-Mn-Nb oxide system *ACS Comb. Sci.* **2017,** *19* (1), 37-46.

15. Stein, H. S.; Jiao, S.; Ludwig, A., Expediting Combinatorial Data Set Analysis by Combining Human and Algorithmic Analysis. *Acs Comb Sci* **2017,** *19* (1), 1-8.

16. Ilavsky, J., Nika: software for two-dimensional data reduction. *J Appl Crystallogr* **2012,** *45*, 324-328.

17. Toby, B. H.; B., V. D. R., GSAS-II: the genesis of amodern open-source all purpose crystallography software package. *Journal of Applied Crystallography* **2013,** *46* (2), 5.

18. Hammersley, A. P., FIT2D: a multi-purpose data reduction, analysis and visualization program. *J Appl Crystallogr* **2016,** *49*, 646-652.

19. Ashiotis, G.; Deschildre, A.; Nawaz, Z.; Wright, J. P.; Karkoulis, D.; Picca, F. E.; Kieffer, J., The fast azimuthal integration Python library: pyFAI. *J Appl Crystallogr* **2015,** *48*, 510-519.

20. Baker, J. L.; Jimison, L. H.; Mannsfeld, S.; Volkman, S.; Yin, S.; Subramanian, V.; Salleo, A.; Alivisatos, A. P.; Toney, M. F., Quantification of Thin Film Crystallographic Orientation Using X-ray Diffraction with an Area Detector. *Langmuir* **2010,** *26* (11), 9146-9151.

21. Hernandez-Rivera, E.; Coleman, S. P.; Tschopp, M. A., Using Similarity Metrics to Quantify Differences in High-Throughput Data Sets: Application to X-ray Diffraction Patterns. *Acs Comb Sci* **2017,** *19* (1), 25-36.

22. Iwasaki, Y.; Kusne, A. G.; Takeuchi, I., Comparison of dissimilarity measures for cluster analysis of X-ray diffraction data from combinatorial libraries. *npj Computational Materials* **2017,** *3* (1), 4.

23. Johnson, D. H., Signal-to-noise ratio. *Scholarpedia* **2006,** *1* (12), 2088.
32